\documentclass[final,5p,times,twocolumn]{elsarticle}

\usepackage{graphics}

\usepackage{amssymb}

\journal{Physics Letters B}

\begin{document}

\begin{frontmatter}



\title{The fission barriers in Actinides and superheavy nuclei in covariant
density functional theory}


\author{S.\ Karatzikos$^1$, A.\ V.\ Afanasjev$^2$, G.\ A.\ Lalazissis$^1$, P.\ Ring$^3$}

\address{$1$ Department of Theoretical Physics, Aristotle University
of Thessaloniki GR-54006, Thessaloniki, Greece}

\address{$2$ Department of Physics and Astronomy, Mississippi State
University, Mississippi State, Mississippi 39762, USA}

\address{$3$ Physik-Department, Technische Universit\"{a}t M\"{u}nchen
D-85747, Garching, Germany}

\begin{abstract}
The impact of pairing correlations on the fission barriers is
investigated in Relativistic Hartree Bogoliubov (RHB) theory and
Relativistic Mean Field (RMF)+BCS calculations. It is concluded that
the constant gap approximation in the usual RMF+BCS calculations does
not provide an adequate description of the barriers. The RHB
calculations show that there is a substantial difference in the
predicted barrier heights between zero-range and finite range pairing
forces even in the case when the pairing strengths of these two
forces are adjusted to the same value of the pairing gap at the
ground state.
\end{abstract}

\begin{keyword}
Fission barrier\sep covariant density functional theory\sep pairing
\PACS 21.60.Jz\sep 24.75.+i\sep 25.85.CA\sep 25.85.-w

\end{keyword}

\end{frontmatter}



\section{Introduction}


A study of the (static) fission-barrier height $B_{f}^{st}$ of nuclei
is motivated by the importance of this quantity for several physical
phenomena. Many heavy nuclei decay by spontaneous fission, and the
size of the fission barrier is a measure of stability of the nucleus
reflected in the spontaneous fission lifetimes of these
nuclei~\cite{SP.07}. The probability for the formation of a
super-heavy nucleus in a heavy-ion-fusion reaction is also directly
connected to the height of its fission barrier~\cite{IOZ.02}. The
height $B_{f}^{st}$ is a decisive quantity in the competition between
neutron evaporation and fission of a compound nucleus in the process
of its cooling. The large sensitivity of the cross section $\sigma$
for the synthesis of the fissioning nuclei on the barrier height
$B_{f}^{st}$ stresses a need for accurate calculations of this value.
For example, a change of $B_{f}^{st}$ by 1 MeV changes the calculated
survival probability of a synthesized nucleus by about one order of
magnitude or even more~\cite{IOZ.02}. The population and survival of
hyper-deformed states at high spin also depends on the fission
barriers, see e.g. Refs.~\cite{DPS.04,AA.08}. In addition, the
$r-$process of stellar nucleosynthesis depends (among other
quantities such as masses and $\beta$-decay rates) on the fission
barriers of very neutron-rich nuclei~\cite{AT.99,MPR.01}.


The fission barriers for Actinides and super-heavy nuclei show
significant differences when calculated by various theoretical
approaches (see, for example, Fig.~25 in Ref.~\cite{SP.07} and Fig.~2
in Ref.~\cite{IOZ.02}). Many extensive calculations of fission
barriers have been done in the framework of the
macroscopic+microscopic (MM) method (see Refs.
\cite{BDJ.72,BPL.81,GSP.99,MSI.04,MS.05,MSI.09} and references
therein). In addition, after the pioneering work of Flocard et
al.~\cite{FQV.74} a growing number of self-consistent investigations
have been reported in recent years. These calculations are based on
modern energy density functionals. After semiclassical investigations
\cite{GHB.80} there are now fully quantum mechanical calculations
available based on
Skyrme~\cite{BRR.98,BRR.00a,BBM.04,SGP.05,SK.06,SDN.07},
Gogny~\cite{BGG.84,WER.02} or RMF~\cite{BMR.94,RMR.95,BRR.00a,AA.08}
functionals. The majority of these calculations are restricted to
axially symmetric shapes. In reality, however, the axially symmetric
fission barriers sometimes indicate only an upper limit, because
calculations including triaxiality~\cite{SDN.07,CDH.96,WER.02,BRR.98}
have found a lowering of the fission barrier by up to several MeV.
This lowering strongly depends on the proton and neutron number and
on the model employed.

So far all investigations of fission barriers based on covariant
density functional theory (CDFT) have been performed in the RMF+BCS
framework~\cite{RMR.95,BRR.98,BBM.04}. One of the major goals of the
current manuscript is to perform the detailed comparative study of
the inner fission barriers in the RMF+BCS and in the relativistic
Hartree-Bogoliubov (RHB) frameworks. In particular, we try to
understand to what extent the properties of the first barrier are
influenced by the impact of different pairing schemes and how the
uncertainties in the extrapolation of the pairing strength towards
super-heavy region affect the fission barriers in these nuclei. In
order to save computational time we do not consider the outer fission
barriers in the current investigation. This restriction has its own
merits. The inner barriers are generally better measured than the
outer ones, and they are certainly more important for the
$r$-process, since they determine the thresholds. Furthermore,
spontaneous fission lifetimes tend to be dominated by the inner
barrier, even if occasionally an outer barrier can have a crucial
effect if it is large enough. The consideration of only inner fission
barriers allows us to restrict the calculations to reflection
symmetric shapes, because it has been found in several
investigations~\cite{RMR.95,SDN.07,SK.06,CDH.96,SGP.05} that
odd-multipole deformations (octupole, etc.) do not play a role in the
inner fission barrier of the Actinides and of superheavy nuclei.

The manuscript is organized as follows. The theoretical framework,
the selection of pairing schemes and the numerical details of
calculations are discussed in Sect.~\ref{Theory}. In
Sect.~\ref{Barrier} we study the influence of pairing correlations on
the shape and the height of the inner fission barrier. Experimental
data on fission barriers in Actinides and superheavy nuclei is
compared with the results of such calculations in
Sect.~\ref{Act-SH-exp}. Finally, Sect.~\ref{Conclusions} contains the
main conclusions of our work.

\section{Theoretical framework and numerical details}
\label{Theory}

The calculations discussed in the current manuscript are based on two
mean field methods to treat pairing correlations in nuclei:

\bigskip

\subsection{RMF+BCS calculations}

The RMF+BCS scheme is rather simple. The RMF-equations are solved and
at each step of the iteration the BCS occupation probabilities
$v_{k}^{2}$ are determined. These quantities are used in the
calculation of densities, energies and new fields for the next step
of the iteration. Two methods are available for the evaluation of the
occupation numbers $v_{k}^{2}$, either the "constant gap"
approximation~\cite{Vau.73,GRT.90} or the solution of the gap
equations based on a seniority force with the strength parameters
$G_{\tau}$ for neutrons ($\tau=n$) and protons ($\tau=p$) (denoted in
the following "constant $G$").

In the first case (constant gap) one starts with fixed gap parameters
$\Delta$ and
uses the BCS expression%
\begin{equation}
v_{k}^2=\frac{1}{2}\left(  1-\frac{\varepsilon_{k}-\lambda}{E_{k}%
}\right)  \label{v2}%
\end{equation}
with $E_{k}=\sqrt{(\varepsilon_{k}-\lambda)^{2}+\Delta^{2}}$ for
the occupation numbers and for the pairing energy%
\begin{equation}
E_{\rm pair}=-\Delta%
\sum\limits_{k>0}
u_{k}v_{k} %
\label{Epair}%
\end{equation}
$\varepsilon_{k}$ are the eigenvalues of the Dirac equation and the
chemical potentials $\lambda$ are determined by the particle numbers.

In the second case (constant $G$) one starts with pairing strengths
parameters $G$ and solves in each step of the iteration the
gap equation~\cite{RS.80}%
\begin{equation}
\frac{1}{G}=%
\sum\limits_{k>0}\frac{1}{2E_{k}}%
\label{gap-equation}%
\end{equation}
and determines the gap parameters
\begin{equation}
\Delta=G\sum\limits_{k>0}u_{k}v_{k}%
\label{delta-BCS}%
\end{equation}
in a self-consistent way. This method is based on the residual
interaction of the seniority model \cite{RS.80} and its strength
parameters $G$.

In principle both methods are equivalent if the constants $G$ are
adjusted properly to the corresponding gap parameters $\Delta$ (or
vise versa). There is, however a difference between the two models
when we carry out calculations along the fission path by constraining
the deformation. In the first case (constant gap) the gap parameters
$\Delta$ are kept fixed as a function of the deformation. This means
the equivalent strength parameters $G$ change with deformation. In
the second case (constant $G$) we use a fixed pairing force with
constant strength parameters $G$ for all deformations and the gap
parameters $\Delta$ change with deformation. This situation is, of
course, closer to reality.

\subsection{Fully self-consistent RHB calculations}

The second method used in this investigation is based on a fully
self-consistent solution of the relativistic Hartree-Bogoliubov (RHB)
equations as introduced in Refs. \cite{KuR.91,Ring.96}: %
\begin{equation}
\left(
\begin{array}
[c]{cc}%
h_{D}-\lambda & \Delta\\
-\Delta^{\ast} & -h_{D}^{\ast}+\lambda
\end{array}
\right)  \left(
\begin{array}
[c]{c}%
U\\
V
\end{array}
\right)  _{k}=E_{k}\left(
\begin{array}
[c]{c}%
U\\
V
\end{array}
\right)  _{k}%
\end{equation}
where $h_{D}$ is the Dirac hamiltonian of RMF-theory and the pairing
field
\begin{equation}
\Delta_{12}=\sum\limits_{3<4}V_{1234}^{\rm pp}\kappa_{34}%
\label{Delta}%
\end{equation}
is determined in a self-consistent way by the pairing tensor
\begin{equation}
\kappa_{12}=\sum_{k}V_{2k}^{\ast}U^{}_{1k}, \label{kappa}
\end{equation}
and an effective pairing interaction $V^{\rm pp}$. The pairing energy
is given by
\begin{equation}
E_{\rm pair}=-\frac{1}{2}{\rm Tr}\Delta\kappa.%
\label{E-pair-HB}
\end{equation}
In the present investigations we compare two different pairing
interactions. First we use the Brink-Booker part of the Gogny force
with finite range
\begin{equation}
V^{\rm pp}(1,2)=\sum_{i=1,2}e^{-(r/{\mu_{i}})^{2}}
(W_{i}~+~B_{i}P^{\sigma}-H_{i}P^{\tau}-M_{i}P^{\sigma}P^{\tau}),
\label{Gogny}%
\end{equation}
where $P^{\sigma}$ and $P^{\tau}$ are the exchange operators for spin
and isospin and the parameters $\mu_{i}$, $W_{i}$, $B_{i}$, $H_{i}$,
and $M_{i}$ $(i=1,2)$ of this force have been carefully adjusted to
the G-matrix calculations in nuclear matter and to the pairing
properties of finite nuclei all over the periodic table (for details
of this fit see Refs.~\cite{DG.80,Cha.07}). We use here the parameter
set D1S~\cite{BGG.84}.

As second example we use a zero range $\delta$-force%
\begin{equation}
V^{\rm pp}(1,2)~=-V_{0}\delta(\mathbf{r}_{1}-\mathbf{r}%
_{2}),%
\label{Vdelta}%
\end{equation}
as it has been used in many non-relativistic HFB
calculations~\cite{SGP.05}. This interaction does no depend on
density and thus it leads to volume pairing.

In RHB-calculations the pairing gap of BCS-theory is replaced by the
pairing field $\Delta_{12}$ in Eq.~(\ref{Delta}). In order to have a
simple measure for the size of pairing correlations in RHB theory we
therefore introduce in the following calculations the "average gap"
in the canonical basis%
\begin{equation}
\langle\Delta\rangle=\frac{\sum\limits_{k}v_{k}^{2}\Delta_{k\bar{k}}}
{\sum\limits_k v_k^2}%
\label{Delta-average}%
\end{equation}
where $v_{k}^{2}$ are the eigenvalues of the density matrix $\rho_{12}%
=\sum_{k}V_{2k}^{\ast}V^{}_{1k}$ in the canonical basis and
$\Delta_{k\bar{k}}$ are the diagonal matrix elements of the pairing
field $\Delta$ in this basis. For details see Ref.~\cite{RS.80}

\subsection{The pairing window}

Pairing is restricted to the vicinity of the Fermi surface and the
size of this vicinity is essentially characterized by the gap
parameter $\Delta$. This parameter determines the distribution of
occupations numbers $v_{k}^{2}$ and therefore most of the
experimental quantities depending on pairing correlations. Of course,
in BCS or HFB theory based on the seniority model or on zero
range forces the sums in Eqs. (\ref{Epair},\ref{gap-equation},\ref{delta-BCS}%
, or \ref{kappa}) show an ultra-violet divergence  and one has to
limit the sums in
Eqs.~(\ref{Epair},\ref{gap-equation},\ref{delta-BCS}) to a pairing
window $\varepsilon_k<E_{\rm cut-off}$ and in Eq.~(\ref{kappa}) to
$E_k<E_{\rm cut-off}$. Concerning this pairing window, there is a
difference between the methods of constant gap and constant $G$. In
the method of constant gap, the essential quantity $\Delta$ is fixed
and determined in one or another way by experiment. The size of the
pairing window enters only in the calculation of the pairing energy
(\ref{Epair}) which is not measurable. For reasonable pairing windows
the pairing energy is small as compared to the binding energy of the
system and by this reason one often finds the remark in the
literature that the results do not depend on the pairing window.
However, the situation is different for the method of constant $G$ or
for zero range forces. In this case the size of the effective pairing
constants $G$ or $V_0$ in Eq.~(\ref{Vdelta}) have to be adjusted in
such a way, that the resulting gap parameters $\Delta$ correspond
roughly to the experimental gaps. As a consequence there is a strong
connection between the strength of the pairing force and the cut-off
energy $E_{\rm cut-off}$. It makes no sense to give only the strength
parameter $G$ (or $V_0$) if the corresponding value of $E_{\rm
cut-off}$ is not known. Essentially one is left with the problem that
one has one experimental quantity $\Delta$ and two unknowns $G$ (or
$V_0$) and $E_{\rm cut-off}$. Usually one chooses a fixed value for
$E_{\rm cut-off}$ in a somewhat arbitrary way and determines the
corresponding strength parameters in such way that the gap parameter
$\Delta$ corresponds more or less to the experimental value.

Bulgac~\cite{Bul.02} has proposed an elegant way to regularize the
divergence of the gap equation in $r$-space by removing the divergent
part of the pairing tensor~(\ref{kappa}) producing in this way a
zero-range pairing force, which depends only on one parameter. On the
other side, as it is clearly seen in momentum space, the two
parameters $G$ (or $V_0$) and $E_{\rm cut-off}$ are connected to two
different physical quantities, namely, to the strength and the range
of the effective pairing force. This means that $E_{\rm cut-off}$
should not be chosen in an arbitrary way, but it should be determined
by the range of the force. Of course, it not clear which experimental
quantity is really sensitive to the range of the effective pairing
force. To search for such quantities one could do calculations with
finite range pairing forces of different range adjusting in each case
the remaining parameters to the experimental gap. In this context we
will study in Sect.~\ref{Barrier} the influence of the pairing window
of zero-range pairing forces on the size of the fission barriers.

Gogny~\cite{DG.80,BGG.91} has avoided the problem of the pairing
window by using a finite range force of Gaussian type where the range
is adjusted to a G-matrix, i.e. via an ab-initio calculation directly
to the properties of the bare nucleon-nucleon interaction. The fact
that this pairing force is extremely successful in reproducing an
astonishing number of experimental data both in
non-relativistic~\cite{DG.80,ER.89,VER.97}  and relativistic energy
density functionals~\cite{GEL.96,AKR.99,VALR.05} shows that this
choice of the range is deeply connected to the physics of the nuclear
many-body problem.

\subsection{Numerical details}

In the current manuscript, we use new versions of the RMF+BCS and RHB
codes which were specifically designed to describe axially symmetric
nuclei with large elongation. They allow to use different numbers of
integration points in the directions along the symmetry axis and in
perpendicular direction thus allowing a better numerical description
of highly elongated systems. As discussed in
Refs.~\cite{GRT.90,RGL.97} the RMF+BCS and RHB equations are solved
in the basis of an axially deformed harmonic oscillator potential
characterized by the deformation parameter $\beta_{0}$ and oscillator
frequency $\hbar \omega_{0}=41A^{-1/3}$ MeV. The truncation of basis
is performed in such a way that all states belonging to the shells up
to $N_{F}$  fermionic shells and $N_{B}$  bosonic shells are taken
into account. The computational time increases considerably with the
increase of $N_{F}$ but it is much less dependent on $N_{B}$. Thus, a
special attention has been paid to the selection of $N_{F}$ of the
basis which would allow for a systematic study of fission barriers in
the nuclei of interest, providing at the same time a reasonable
numerical accuracy in the predictions of the physical observables.

Extensive tests of numerical convergence have been performed in the
spherical, normal-deformed and superdeformed ($\beta_{2}\sim0.7-1.0$)
minima in the RMF calculations without pairing on the example of the
nuclei $^{238}$U and $^{304}120$ with $Z=120$ and $N=184$. Contrary
to the previous studies of the convergence in the RMF framework which
were based on the comparison of the $N_{F}$ and $N_{F}+2$ results, we
first defined the \textquotedblleft exact\textquotedblright\ solution
(extending the calculations up to $N_{F}=36,N_{B}=36$) which does not
change with the increase of $N_{F}$, and then found the truncation
scheme for a basis which provides sufficient numerical accuracy. It
turns out that the binding energies for $N_{F}=20$ and $N_{B}=20$ are
described with an accuracy of 200 keV as compared with the exact
solution. It was also checked that the inner fission barrier is
described with an accuracy of $\sim100$ keV both in the RMF+BCS and
RHB calculations in this truncation scheme. Thus all the following
calculations are performed with $N_{F}=N_{B}=20$.

\section{The role of pairing in defining the inner fission barrier}%
\label{Barrier}

\subsection{Comparison of different pairing schemes}

The height of the fission barrier depends on the pairing. This is a
well known fact from the early work of Ref.~\cite{SS.68} where it was
shown for reflection symmetric shapes that barrier heights decrease
with increasing nuclear pairing. The study of the fission barrier in
$^{240}$Pu within the RMF+BCS framework of Ref.~\cite{RBR.99} showed
that the enhancement of the pairing strength of the zero-range
pairing force by 20\% decreases the fission barrier by approximately
2 MeV. Similar results were obtained also in non-relativistic HFB
calculations based on the Skyrme forces in Ref.~\cite{SGP.05}: the
reduction of the pairing strength of zero-range pairing forces by
approximately 15\% leads to an increase of the fission barriers by
roughly 2 MeV in the Actinides and can enhance it by up to 4 MeV in
superheavy nuclei thus doubling the barrier heights. The impact of
pairing on fission barriers has also been studied in the
Skyrme+HF+BCS approach using a seniority pairing force and zero-range
$\delta$-interactions with different forms of density dependence in
Ref.~\cite{SDN.07}.

\begin{figure}[t]
\includegraphics[angle=0,width=8.5cm]{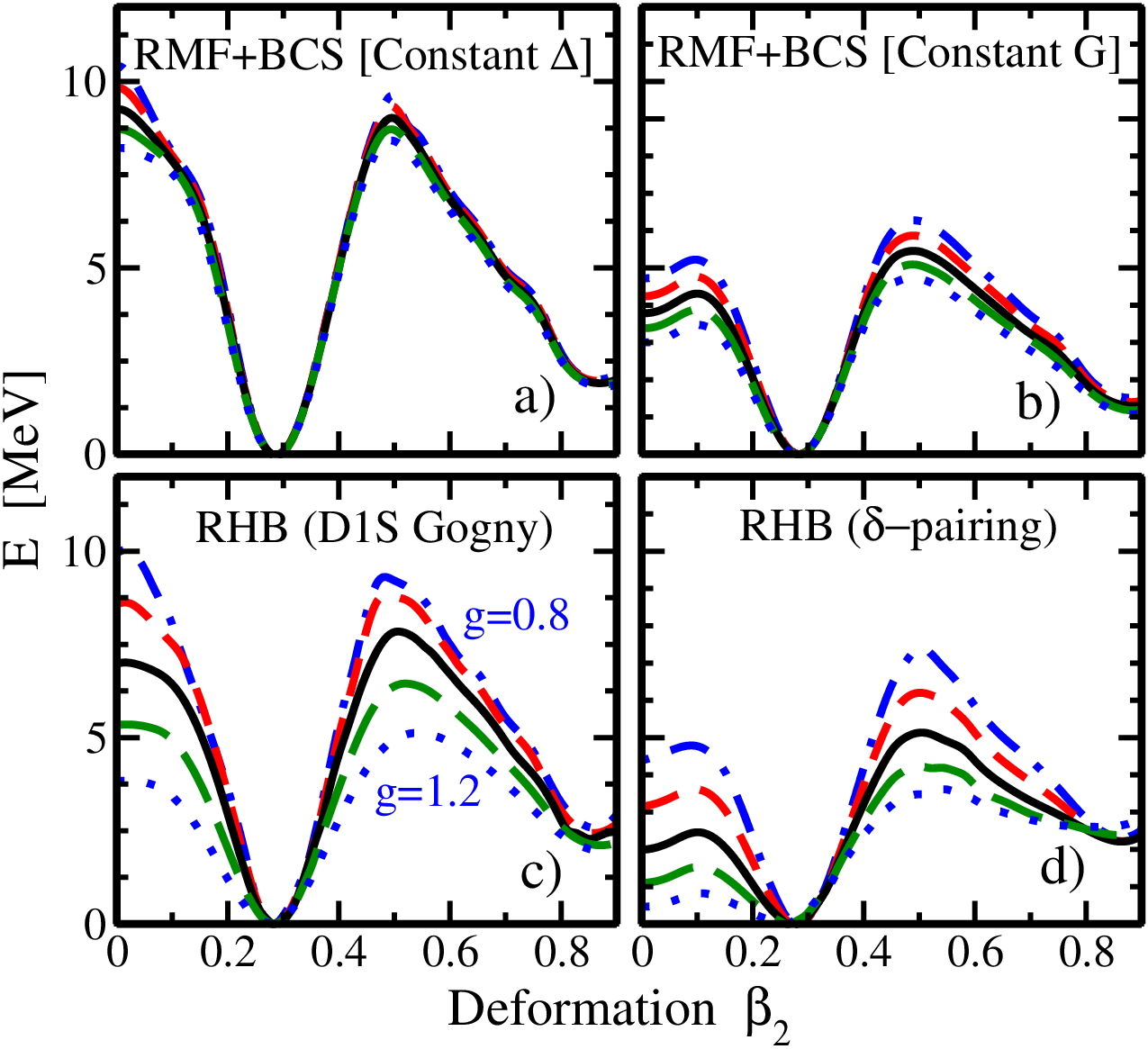}%
\vspace{0.0cm}%
\caption{(Color online) Potential energy surfaces in $^{240}$Pu
obtained in different pairing schemes with the NL3 parameterization
of the RMF Lagrangian and different values of the scaling factor $g$.
They are normalized to the energy of the normal-deformed (ND)
minimum.
Further details are given in the text.}%
\label{fig1}%
\end{figure}

In Fig.~\ref{fig1} we investigate as a typical example the inner
fission barrier of the nucleus $^{240}$Pu first by RMF+BCS and then
by full RHB calculations with four different pairing schemes. The
major goal of this figure is to show the dependence of the potential
energy surface (PES) and its profile on the pairing strength as a
function of the deformation. Therefore we introduce a scaling factor
$g$ which allows us to change the strength of the pairing
correlations in an appropriate way. Details will be discussed below.
Note that we have not tried to adjust the pairing strengths in all
these pairing schemes. As a result, the PES of the different panels
of Fig.~\ref{fig1} cannot be compared directly.

\begin{figure*}[t]
\includegraphics[angle=0,width=15.5cm]{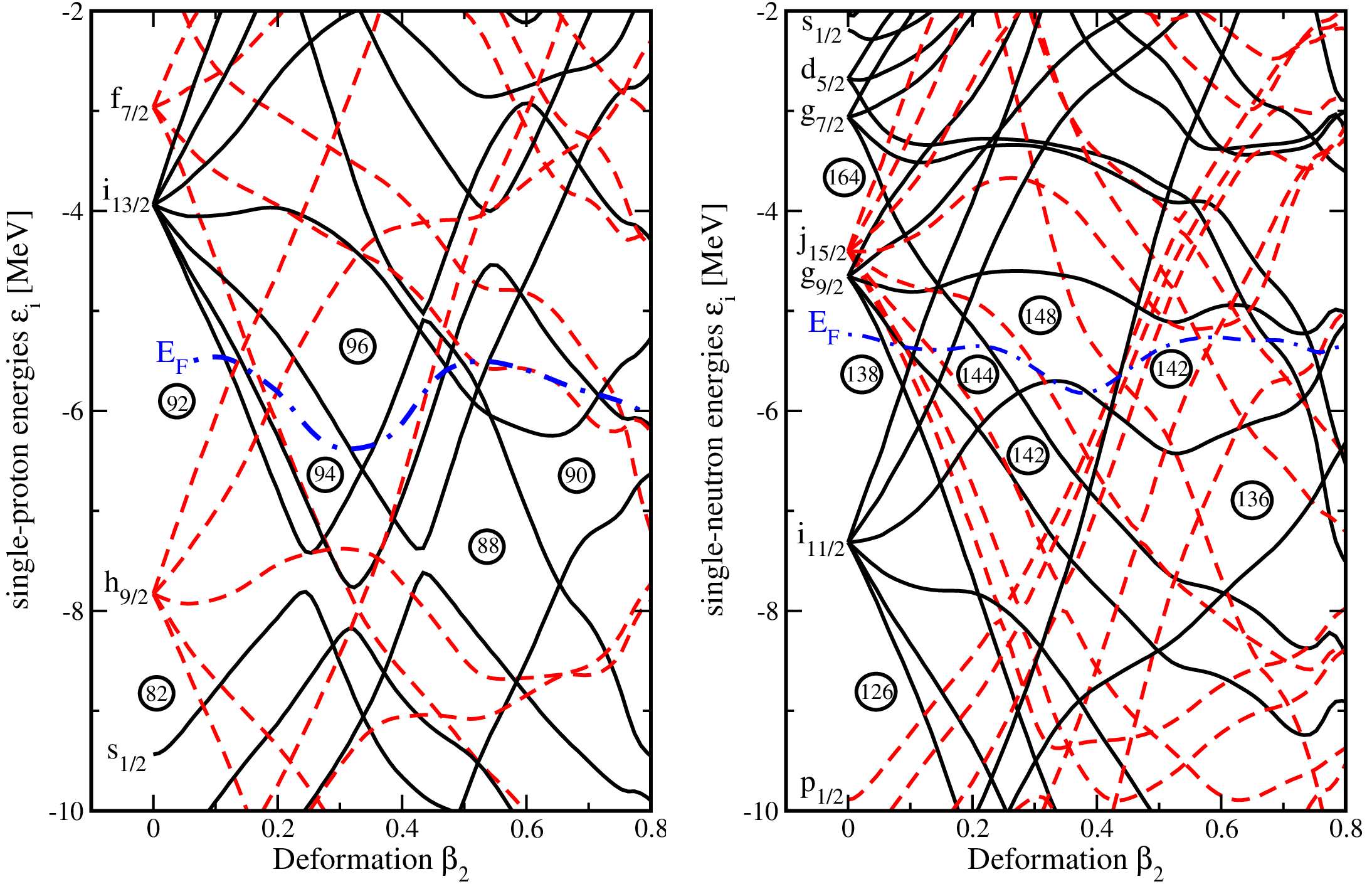}%
\vspace{0.0cm}%
\caption{Proton and neutron single-particle energies in $^{240}$Pu as
a function of the quadrupole deformation $\beta_{2}$. They are
obtained in the RMF+BCS calculations with constant gap approximation
and the NL3 parameterization of the RMF Lagrangian. Solid and dashed
lines are used for positive and negative
parity states, respectively.}%
\label{fig2}%
\end{figure*}

Figs.~\ref{fig1}a and b show RMF+BCS calculations with constant gap
and with constant $G$ within a pairing window of $E_{\rm cut-off}=60$
MeV. In the first case shown in Fig.~\ref{fig1}a we use the same
constant gap parameters $\Delta_\tau$ for the entire deformation
range. The size of this pairing gap parameters is determined for the
ground state in the first minimum by a prescription given in
Ref.~\cite{MN.92}
\begin{equation}
\Delta_{n}=\frac{4.8}{N^{1/3}}\quad\mathrm{MeV}, \ \ \ \ \Delta
_{p}=\frac{4.8}{Z^{1/3}}\quad\mathrm{MeV}.%
\label{constant_gap}%
\end{equation}
In order to see how the barrier depends in this case on the gap
parameter we multiply this choice for the gap parameter with a
scaling factor $g$
\begin{equation}
\Delta_\tau \rightarrow g\Delta_\tau
\label{scaling-Delta}%
\end{equation}
in the range between $g=0.8$ and $g=1.2$ in a step of 0.1. The
different potential energy surfaces obtained in this way are
normalized at the ground state. We observe only small changes of the
barrier for the different values of $g$. The height of the barrier is
the largest for the reduced gap with $g=0.8$ and it drops only
slightly with increasing pairing correlations.

In Fig.~\ref{fig1}b we show results of similar calculations, but now
with constant $G$. We start in the first minimum with the same values
for the gap parameters as in Fig.~\ref{fig1}a and determine in a
self-consistent way for each value of $g$ the strength parameters
\begin{equation}
G_\tau = \Delta_\tau/\sum_k u_kv_k
\label{G}%
\end{equation}
In a calculation with constant $G$ this choice leads in the first
minimum of Fig.~\ref{fig1}b to the same gap parameters $\Delta_\tau$
as in Fig.~\ref{fig1}a. Of course, we also could have scaled the
strength parameters $G_\tau\rightarrow g G_\tau$ directly. However,
because of the strong non-linearity of the gap equation this leads to
dramatic changes in the potential energy surfaces and the two
calculations with constant gap and constant $G$ would have been hard
to compare. Therefore in Fig.~\ref{fig1}b we compare the potential
energy surfaces obtained from constant $G$-calculations using
$G_\tau$-values derived according to Eq.~(\ref{G}) after a
self-consistent solution for the various scaled gap parameters. In
this case the calculations for the various $g$-values in the first
minimum are identical in Fig.~\ref{fig1}a and Fig.~\ref{fig1}b.
However, the remaining part of the energy surfaces are very
different. In particular the height of the barrier is now reduced by
nearly a factor 2 and the relative changes for the various $g$-values
are larger than in the case of constant gap.

This behavior can be understood by the fact that the minima and
maxima (saddle points) along the fission path are induced by shell
effects. The microscopic mechanism of this process is related to the
density of the single-particle states in the vicinity of the Fermi
level which is a function of the deformation~\cite{Str.67,Str.68}. In
Fig.~\ref{fig2} we show Nilsson diagrams for protons and neutrons in
the nucleus $^{240}$Pu. Minima in the potential energy surfaces are
stable configurations. They corresponds to a region of low level
density, whereas a saddle point occurs in the vicinity of level
crossings, a region of higher level density. In regions of high level
density it is easier for the quasi-particles to spread around the
Fermi surface and therefore the size of the pairing correlations
depends strongly on the level density. As a consequence we find a
relatively small pairing gap at the minima of the PES and large
pairing at the saddle points. Of course, this effect is restricted to
the cases where we determine the gap in a self-consistent way using a
pairing interaction, which does not change along the fission path.

\begin{table}[b]
\caption{Gap parameters for neutron and protons for several
deformations along the fission path obtained by calculations with
constant $G_n=0.0399$ MeV and $G_p=0.0616$ MeV, i.e. for the full
black curve in Fig.~\ref{fig1}b.}%
\label{tab1}
\renewcommand{\tabcolsep}{0.5pc}
\renewcommand{\arraystretch}{1.4}
\begin{tabular}
[c]{lcccc}%
\hline\hline
           & $\beta=0$ & ND minimum & saddle point & SD minimum\\
$\Delta_n$ & 1.560 & 0.912 & 1.335 & 1.033\\
$\Delta_p$ & 1.280 & 1.056 & 1.535 & 1.236\\
\hline\hline
\end{tabular}
\\[2pt]
\end{table}%

Pairing correlations with partial occupations of the different levels
in the neighborhood of the Fermi surface have the tendency to wash
out shell effects and by this general argument we expect a reduction
of the barrier height with increasing pairing correlations. In fact,
we observe this tendency in all four panels of Fig.~\ref{fig1}.
However the gap at the Fermi surface is relatively small in nuclei
(roughly 1 MeV) as compared to the shell effects which are of the
order of 1$\hslash\omega_{0}$ ($\approx$5.6 MeV in $^{240}$Pu). Thus
we can understand that in the case of constant gap in
Fig.~\ref{fig1}a a variation of the scaling factor $g$ in the range
of 20\% has almost no impact on the PES. This is a new result not
available in the literature. Note that the constant gap approximation
has been used in early RMF studies
(Refs.~\cite{BMR.94,RMR.95,RTC.02}). In view of the results of the
present analysis such investigations have to be treated with care. On
the other hand the calculations with constant G shown in
Fig.~\ref{fig1}b lead to oscillations of the gap parameters along the
fission path (see Table~\ref{tab1}).

\begin{figure}[th]
\includegraphics[angle=0,width=8.0cm]{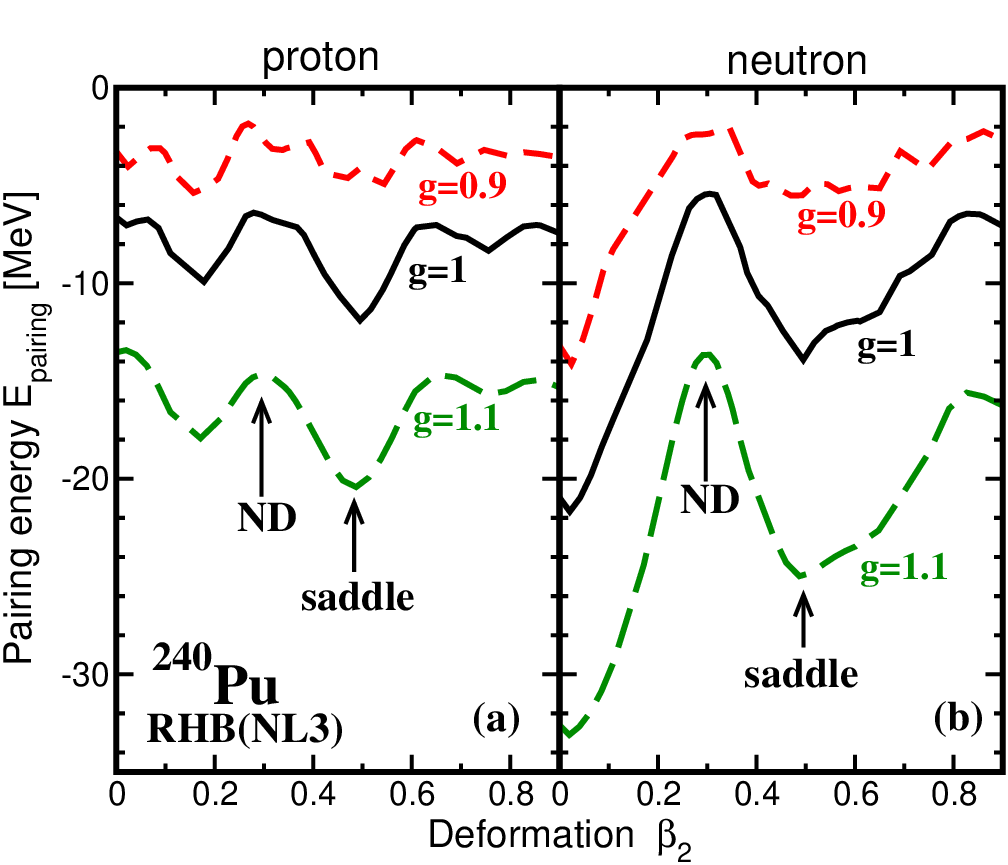}%
\caption{(Color online) Proton and neutron pairing energies in
$^{240}$Pu as functions of $\beta_{2}$-deformation. They are obtained
in the RHB calculations with the Gogny D1S force for indicated values
of $g$. The arrows
show the positions of the ND minimum and saddle point.}%
\label{fig3}%
\end{figure}

The dramatic change of the fission barrier heights is not caused in
the first place by the change of pairing correlations in total, but
rather by the fact that the pairing correlations change along the
fission path. To elaborate more on this oscillating behavior of
pairing correlations we show in Fig.~\ref{fig3} the pairing energies
$E_{\rm pair}$ defined in Eq.~(\ref{E-pair-HB}) for self-consistent
calculations with the finite range Gogny-force D1S in the pairing
channel. We find that these quantities are larger at the saddle point
than at the normal deformed (ND) minimum. With increasing $g$, the
magnitudes of the pairing energies at the saddle point grow faster
than the ones at ND minimum. As a consequence, additional binding due
to pairing (which is related to the pairing energies in a highly
non-linear way) grows faster at the saddle point than at ND minimum
with increasing $g$ and as a result, the fission barrier becomes
lower with increasing $g$. Note, however, that the dramatic changes
in the pairing energy cannot be seen directly in the change of the
barrier height, because they are compensated to some extent by the
fact that larger pairing causes a wider distribution of the
occupation probabilities $v^2_k$ around the Fermi surface.

In Figs.~\ref{fig1}c and~\ref{fig1}d we show potential energy
surfaces obtained from self-consistent RHB calculations based on the
parameter set NL3 using the Gogny D1S of Eq.~(\ref{Gogny}) and the
zero range $\delta$-force of Eq.~(\ref{Vdelta}) in the pairing
channel. For the $\delta$-force we use the strength parameter
$V_0=300$ MeV$\cdot$fm$^{3}$ and a pairing window of $E_{\rm
cut-off}=60$ MeV. In both cases we introduce a scaling parameter of
the pairing interaction
\begin{equation} V_{{}}^{\rm pp}(1,2)\rightarrow
gV_{{}}^{\rm pp}(1,2)
\label{scaling-V}%
\end{equation}
and compare in this way the influence of the strength of the pairing
force on the fission barrier. In both cases we observe a considerable
reduction of the barrier height with increasing $g$-values. This has
the same origin as in the case of the constant $G$ calculations in
Fig.~\ref{fig1}b. Note, however, that we cannot compare directly
these results with those of Fig.~\ref{fig1}b, because we do not scale
the strength parameters $G_\tau$ of the seniority force by the factor
$g$ but rather the resulting gap parameters $\Delta_\tau$ of
Fig.~\ref{fig1}a. This leads to much smaller changes of the
parameters $G_\tau$.

Summarizing the results of this section we observe for all pairing
models in Fig.~\ref{fig1} variations of the PES with pairing.
Increasing pairing leads to a reduction of the barrier height. This
effect is, however, relatively small for calculations with constant
gap. On the other hand for self-consistent calculations the
oscillations in the level density at the Fermi surface induce
pronounced changes of pairing along the fission path and a
considerable reduction of the barrier heights with increasing
strength of the effective interaction in the pairing channel.

\subsection{Finite versus zero range pairing}

The basic advantage of the Gogny force is its finite range, which
automatically guarantees a proper cut-off in momentum
space~\cite{D1S}. Calculations with interactions of finite range
require  a substantial numerical effort and therefore many modern
energy density functionals like Skyrme functionals are based on
forces with zero range. In the $ph$-channel, where either only  the
levels up to the Fermi surface are occupied or where the analytical
form of the occupation numbers $v_{k}^{2}$ in Eq.~(\ref{v2}) leads to
a fast convergence in momentum space, a gradient expansion of finite
range interactions leading to zero-range forces of the Skyrme type
\cite{NV.73} are well justified and successful. This is no longer
true in the $pp$-channel, where the analytical form of the factors
$u_{k}v_{k}$ in the pairing tensor $\kappa$ includes high momenta and
leads to a ultra-violet divergence. The same is true for the
seniority pairing which can be considered as the $J=0$ part of the
surface delta interaction \cite{RS.80}. In all these cases one is
forced to use a cut-off energy $E_{\rm cut-off}$ in order to avoid
divergencies.

\begin{figure}[t]
\includegraphics[angle=0,width=8.0cm]{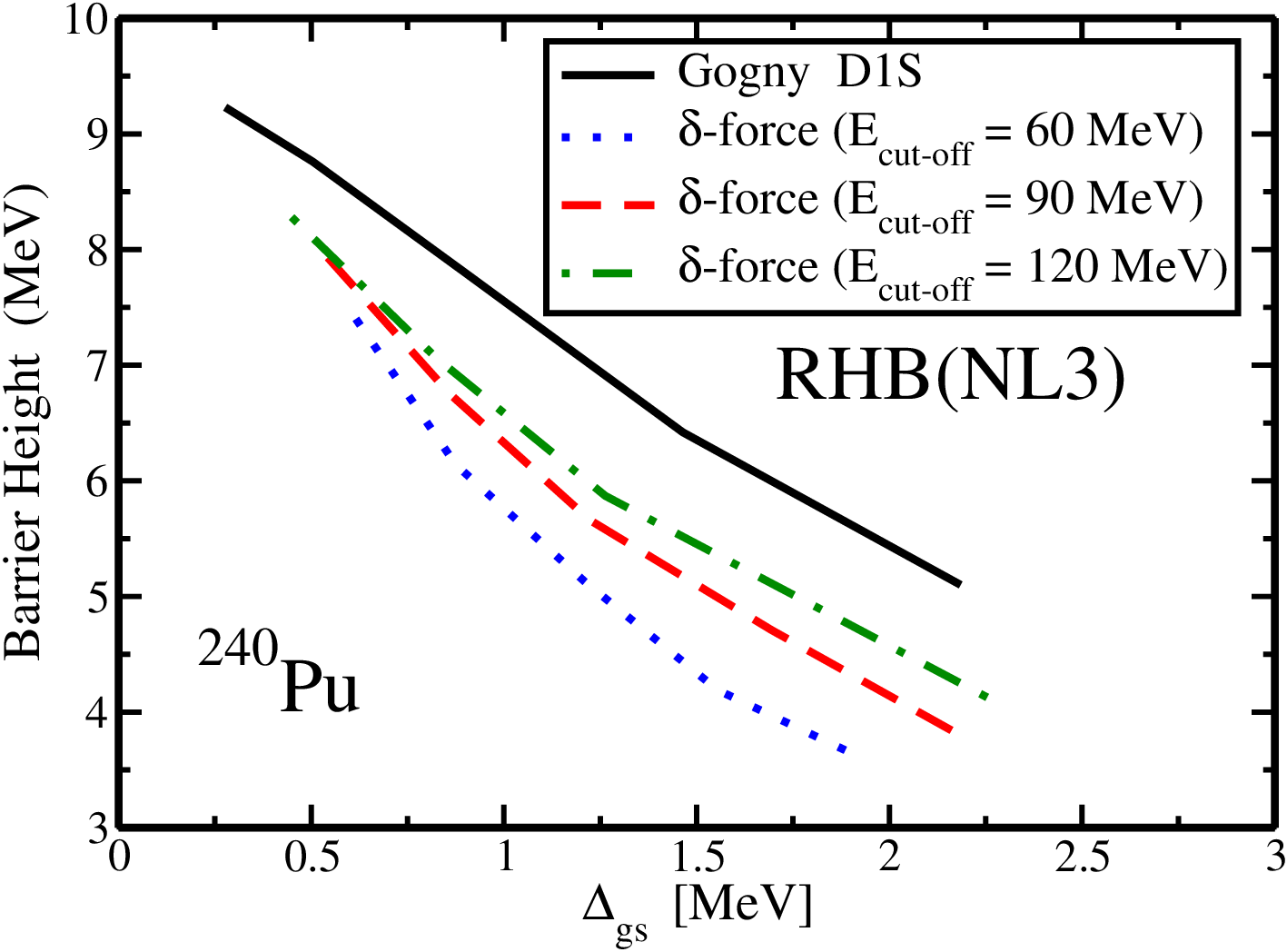}%
\vspace {0.0cm}%
\caption{(Color online) The dependence of the fission barrier height
on the pairing gap in the ground state shown for the RHB calculations
with Gogny D1S force and $\delta$-force. The results for
$\delta$-force are displayed for
3 different values of the cut-off energy $E_{\rm cut-off}$.}%
\label{fig4}%
\end{figure}

The strength parameters of the pairing force $V_0$ (or $G$) are
usually deduced for a fixed value of the cut-off energy $E_{\rm
cut-off}$ from the pairing gaps $\Delta$ extracted from experimental
data such as the odd-even mass differences at the ground state in the
ND minimum and it is assumed that the strength parameters do not
depend on the nuclear deformation. Thus, it is interesting to see how
the description of the fission barriers differs in the RHB
calculations with zero range and with finite range pairing forces
under the condition that the pairing strength parameters are defined
by the same set of experimental data at the normal-deformed minimum.
In particular we will investigate whether the barrier depends on the
choice of the cut-off energy in the case of zero range pairing
forces.

In Fig.~\ref{fig4} we study the dependence of the height of the
fission barrier in the nucleus $^{240}$Pu on the size of pairing
correlations in the ground state of the ND minimum. For this purpose
we carry out RHB calculations with the Gogny force D1S and a
zero-range $\delta$-force in the pairing channel for various values
of the scaling parameter $g$ in Eq.~(\ref{scaling-V}). The resulting
barrier height is plotted as a function of the resulting average
pairing gap (\ref{Delta-average}) in the ground state $\Delta_{\rm
gs}=\frac{1}{2}(\langle\Delta\rangle_{n}+\langle\Delta \rangle_{p})$.
The full (black) curve shows the results obtained with the Gogny
force D1S and the other three lines correspond to calculations with
the zero range $\delta$-force in Eq.~(\ref{Vdelta}) using different
cut-off energies $E_{\rm cut-off}=60,$ $90,$ and 120 MeV. This figure
clearly indicates that there is a strong dependence of the fission
barrier height on the treatment of pairing correlations. For the
Gogny force D1S we find a nearly linear decrease of the fission
barrier height by roughly 1.00 MeV per 0.4 MeV change in the gap
$\Delta_{\rm gs}$. For the same value of $\Delta_{\rm gs}$ the
fission barrier is smaller in the calculations with zero range force
as compared with the ones based on the D1S Gogny force. Again we see
a decreasing barrier height with increasing $\Delta_{\rm gs}$.
However, in the case of zero range force the results also depend on
the cut-off energy $E_{\rm cut-off}$. Only for very small pairing
correlations $\Delta_{\rm gs}\approx0.5$ MeV we have have more or
less the same barrier heights in the calculations with zero range.
For larger pairing we observe increasing differences between barrier
heights calculated with the various values of $E_{\rm cut-off}$. {

\begin{figure}[t]
\vspace{0.0cm}%
\includegraphics[angle=0,width=8.0cm]{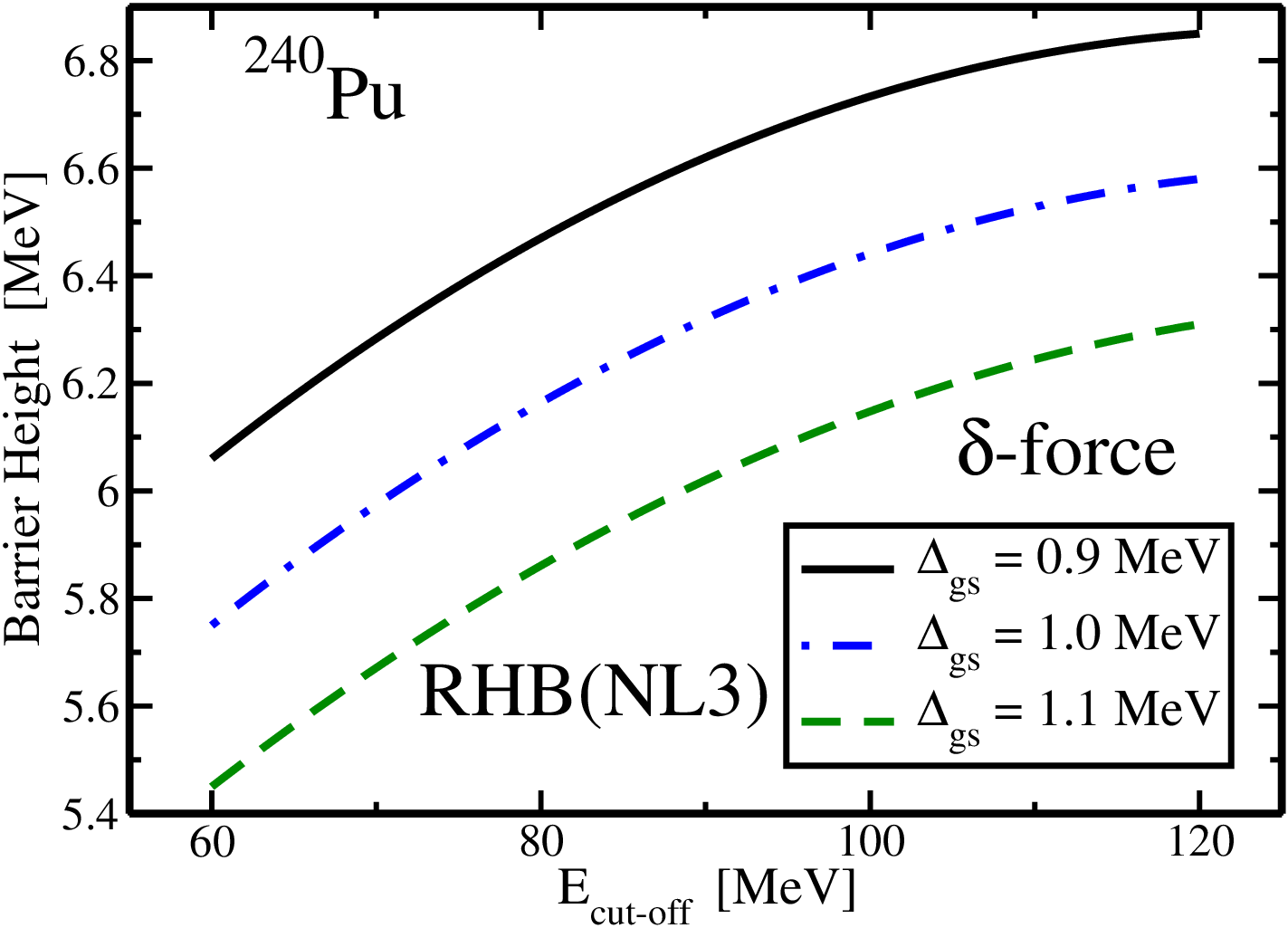}%
\vspace{0.0cm}%
\caption{(Color online) The dependence of the fission barrier on the
cut-off energy $E_{\rm cut-off}$ in the RHB calculations with
$\delta$-force given for 3 different values of pairing gap
$\Delta_{\rm gs}$ in the
normal-deformed ground state.}%
\label{fig5}%
\end{figure}

In particular, for small cut-off energies we find considerably lower
fission barriers than those obtained with the finite range Gogny
force.  This is shown in detail in Fig.~\ref{fig5} where for the same
calculations the height of the fission barrier is plotted as a
function of the cut-off energy $E_{\rm cut-off}$ for three values of
the gap parameter in the ground state ($\Delta_{\rm gs}=0.9,$ $1.0,$
and $1.1$). The dependence of the barrier height on the cut-off
energy is not eliminated even for these large values of $E_{\rm
cut-off}$ which are much larger than the typical ones used in many
calculations with $\delta$-force (see, for example,
Ref.~\cite{SGP.05}).

\subsection{Extrapolation to superheavy nuclei}

  Systematic experimental spectroscopic data, such as odd-even mass
differences and the moments of inertia, which allow to extract the
information on the strength of pairing correlations are available
only up to the proton number $Z\approx102$ and neutron number
$N\approx158$. With increasing proton and neutron numbers such data
become scarce and less reliable. Thus, existing predictions for
superheavy nuclei centered around $Z=120$ and 126 are based on
drastic extrapolations involving the changes of proton number by more
than 20 particles and neutron numbers by 14-26 particles. As with any
extrapolation there is a considerable degree of uncertainty, and
modifications of the strength of the pairing interaction by $\pm$10\%
cannot be ruled out. Indeed, while providing the average description
of pairing properties modern pairing forces show sometimes
appreciable local deviations from experiment for physical observables
which are strongly affected by pairing (see, for example, Fig.~1 in
Ref.~\cite{RBR.99} and Ref.~\cite{HBG.02}).

\begin{figure}[t]
\includegraphics[angle=0,width=8.0cm]{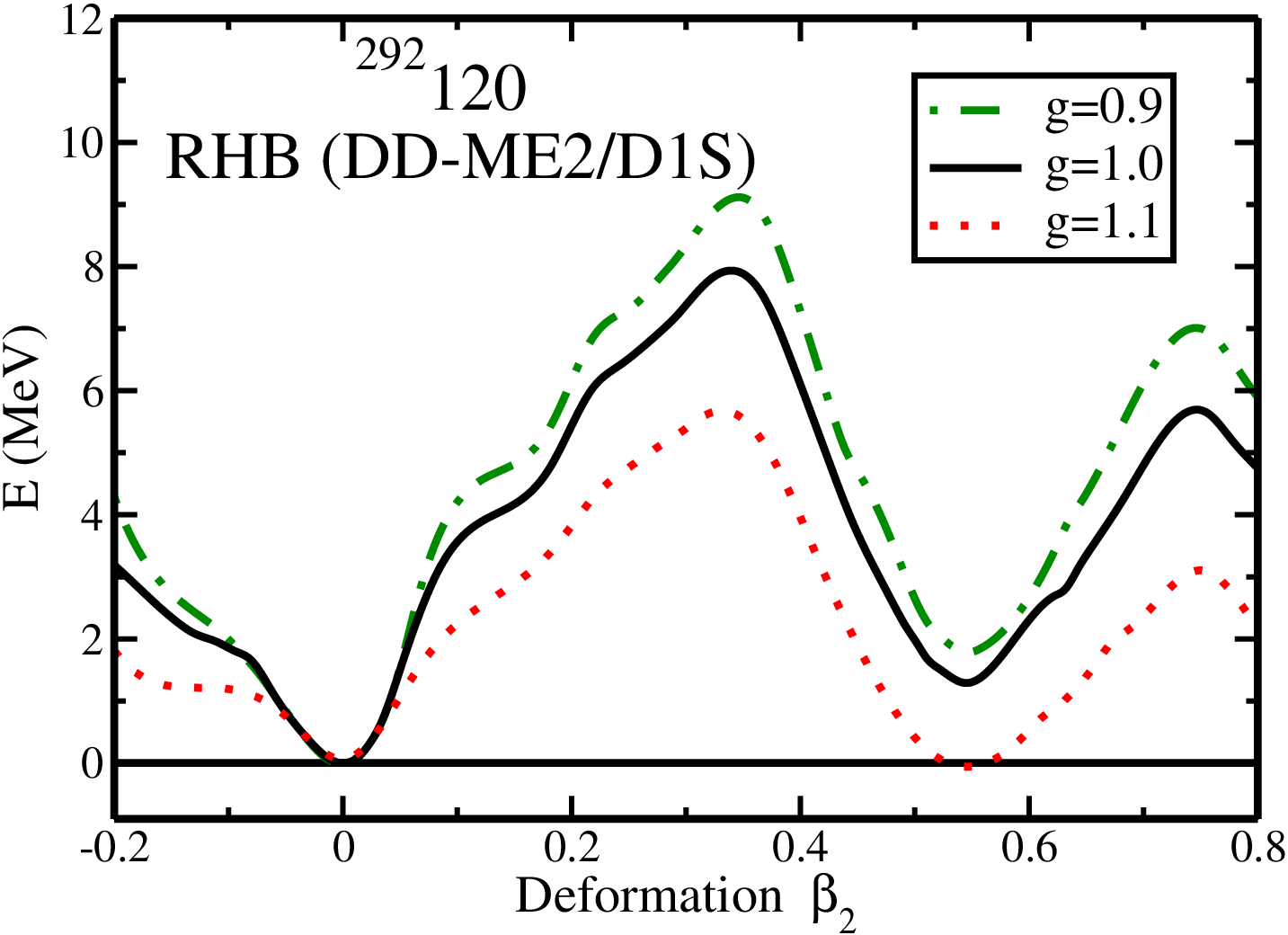}%
\vspace{0.0cm}%
\caption{(Color online) Potential energy surfaces in the $Z=120,
N=172$ nucleus obtained in the RHB calculations with the DD-ME2
parameterization of the RMF Lagrangian
and different values of the scaling factor $g$.}%
\label{fig6}%
\end{figure}

Fig.~\ref{fig6} shows that modifications of the strength of the
pairing interaction by $\pm$10\% have a profound effect on properties
of superheavy nuclei. For example, a decrease of this strength by
10\% (the $g=0.9$ curve in Fig.~\ref{fig6}) increases the inner
fission barrier by $\approx1.0$ MeV, and thus increases the stability
of this nucleus against fission considerably. On the contrary, the
increase of the strength of the pairing force by $\sim$10\% will
modify the situation drastically (the $g=1.1$ curve in
Fig.~\ref{fig6}). Indeed, we find in this case that the superdeformed
minimum becomes lower than the spherical minimum. This could indicate
that the spherical minimum would no longer be the ground state of the
system since the superdeformed minimum is lower in energy. However,
one should remember that odd-multipole deformations (not included in
the current calculations) are important at the outer fission barrier,
and thus if included they will most likely eliminate this
barrier~\cite{BBM.04}. As a consequence the second minimum will not
really stay as a minimum in the PES. In this situation, the ground
state in the spherical minimum is stabilized only by the inner
fission barrier of the $\approx$ 5 MeV height.

\section{Fission barriers in Actinides and superheavy nuclei}
\label{Act-SH-exp}

\begin{figure}[t]
\includegraphics[angle=0,width=8cm]{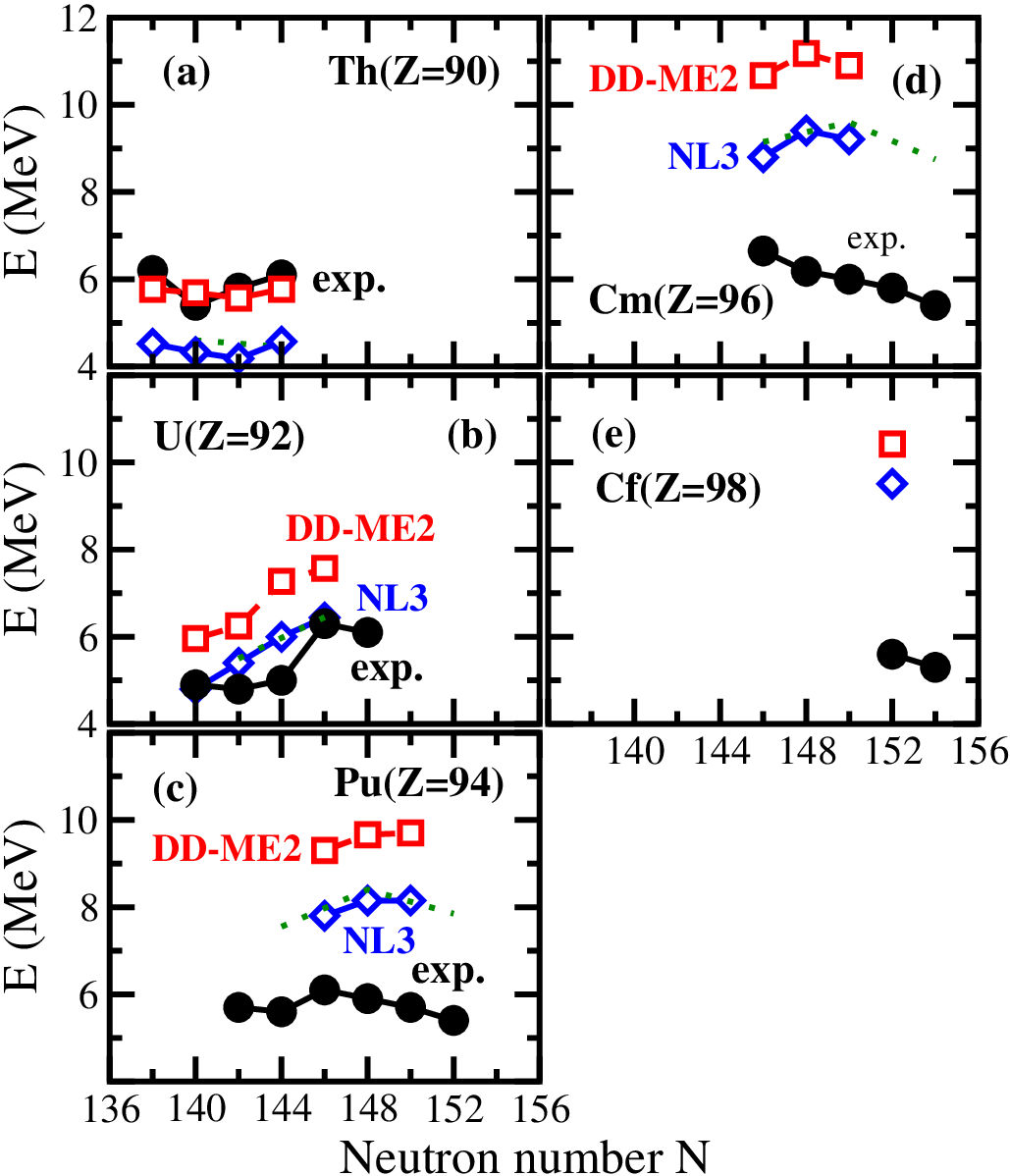}%
\caption{(Color online) Experimental and calculated heights of the
inner fission barriers in Actinide nuclei. Experimental data are
taken from Table IV in Ref.~\cite{SGP.05}. A typical uncertainty in
the experimental values, as suggested by the differences among
various compilations, is of the order of $\pm0.5$ MeV~\cite{SGP.05}.
Results of RHB calculations with the parameter sets NL3 and DD-ME2 of
the RMF Lagrangian are presented. The dotted lines show the results
of the RMF+BCS calculations of Ref.~\cite{BBM.04} which are performed
in steps of $\Delta N=4$.}%
\label{fig7}%
\end{figure}

In Figs.~\ref{fig7} and \ref{fig8} we compare the results of RHB
calculations with the parameter sets NL3~\cite{NL3} and
DD-ME2~\cite{DD-ME2} of the RMF Lagrangian with available
experimental data on the heights of the inner fission barrier in
Actinides and superheavy nuclei. In these calculations the original
Gogny force is used (scaling factor $g=1.0$). It was tested that with
this value of $g$ the moments of inertia of even-even nuclei, which
are very sensitive to the strength of pairing correlations, are
reasonably well described in cranked RHB calculations with the NL3
force in the Nobelium region~\cite{AKF.03} and in the lighter
Actinide nuclei~\cite{A-unpub}.

Fig.~\ref{fig7} compares the calculated heights $E$ of the inner
barrier with experiments for Actinide nuclei. While the dependence of
$E$ on the neutron number $N$ is reasonably well described in both
parameterizations, the dependence on the proton number $Z$ is not
reproduced. The calculated values of $E$ increase with increasing $Z$
while the experimental inner barriers almost do not depend on $Z$.
The inner fission barriers obtained in the calculations with the
DD-ME2 parameterization exceed the ones obtained with the NL3
parameterization by 1-2 MeV.

\begin{figure}[t]
\includegraphics[angle=0,width=8.0cm]{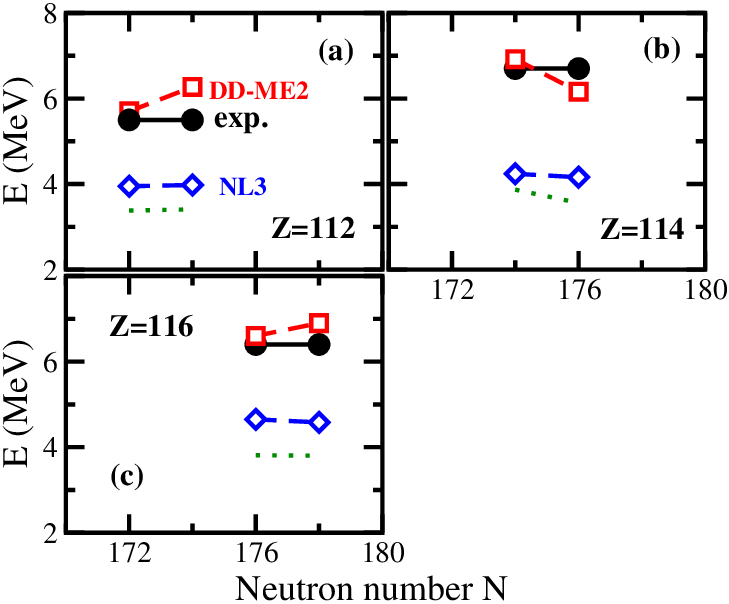}%
\caption{(Color online) The same as in Fig.~\ref{fig7}, but for
superheavy nuclei. Experimental data, which specifies the lower limit
of the heights of fission barriers, are taken from
Ref.~\cite{IOZ.02}. The assignment of the neutron number has some
uncertainty, therefore, the same experimental barrier
appears for the nuclei with the same proton number $Z$.}%
\label{fig8}%
\end{figure}

Fig.~\ref{fig8} clearly shows that the fission barriers in superheavy
nuclei are well described in the RHB calculations with the DD-ME2
parameterization. The calculated barrier heights are either close to
the experimental lower limit or slightly exceed it. Only in the case
of the $(Z=114, N=176)$ system, the calculated fission barrier is
slightly below the experimental value. However, considering the
expected error bars on the height of the fission barrier this fact is
not important. On the contrary, the results of the RHB calculations
with the NL3 force considerably (by $\sim2$ MeV) underestimate the
height of the fission barriers.

The RHB results obtained with the NL3 force are very close to the
ones obtained earlier in the RMF+BCS framework with the same NL3
force in Ref.~\cite{BBM.04}, see Fig.~\ref{fig7} and Fig.~\ref{fig8}.
They are typically within 0.5 MeV of each other. The RMF+BCS barriers
of Ref.~\cite{BBM.04} are either very close to the RHB barriers or
slightly higher (by several hundreds keV) in the Actinide nuclei. On
the contrary, they are lower by 0.5-0.8 MeV than the RHB barriers in
superheavy nuclei. Thus, in going from Actinides to superheavy
nuclei, the change in the height of the fission barriers is more
pronounced (by $\approx 1$ MeV) in the RMF+BCS framework as compared
with the RHB.

Based on the RHB results, one can conclude that the calculations
based on the NL3 parameterization do not leave the room for
triaxiality and the correlations beyond mean field; this conclusion
is consistent with the one obtained in the RMF+BCS framework
earlier~\cite{BBM.04}. The RHB+NL3 calculations underestimate the
experimental fission barriers in some nuclei already on the mean
field level. As discussed in the introduction, the triaxiality can
lower the fission barriers by up to 3 MeV or sometimes even by larger
amount. In addition, the correlations beyond mean field can lower
fission barrier by as much as 1 MeV~\cite{BBM.04} (see also
Ref.~\cite{GSP.99}). It is reasonable to expect that with these
degrees of freedom included the majority of the results with the NL3
force will fall below the experiment. On the contrary, the
experimental fission barriers are overestimated in the RHB
calculations with DD-ME2 force which indicates that there is
sufficient room for triaxiality and the correlations beyond mean
field. It is also interesting to mention that the heights of fission
barriers as obtained with the density-dependent DD-ME2 force are
close to the typical values obtained in the calculations with Skyrme
force~\cite{BBM.04}. On the contrary, the NL3 parameterization of the
RMF Lagrangian, which has no density dependence in the isovector
channel, systematically predicts lower barriers than most Skyrme
forces~\cite{BBM.04}.

\begin{table}[t]
\caption{Excitation energies $E$ (in MeV) of fission isomers as
obtained in the RHB calculations with the Gogny D1S force and
indicated parameterizations of the RMF Lagrangian. Last column shows
experimental data taken from
Ref.~\cite{SZF.02}.}%
\label{tab2}
\renewcommand{\tabcolsep}{0.5pc}
\renewcommand{\arraystretch}{1.4}
\begin{tabular}
[c]{cccc}\hline Nucleus & E(NL3) & E(DD-ME2) & E(exp)\\\hline
$^{236}$U & 1.72 & 2.24 & 2.75\\
$^{238}$U & 1.70 & 1.54 & 2.557\\
$^{240}$Pu & 2.29 & 2.21 & $\sim2.8$\\\hline
\end{tabular}
\\[2pt]\end{table}

An additional constraint on the selection of the force may be
provided by the energy of the 0$^{+}$ fission isomeric state. It
turns out that for Actinides this energy is less dependent on the
variation of the pairing strength than the height of the fission
barrier (see Fig.~\ref{fig1}) thus providing a more robust probe.
Reliable values for the energies of the fission isomers in even-even
nuclei are available only for three nuclei (see Table~\ref{tab2}). It
turns out that in average the NL3 and DD-ME2 parameterizations of the
RMF Lagrangian provide similar results for the energies of the
fission isomers: these energies are underestimated by at least 0.5
MeV.

\section{Conclusions}
\label{Conclusions}

The current investigation of fission barriers within covariant
density functional theory clearly indicates that when aiming at a
quantitative understanding of fission properties of heavy and
superheavy elements, it is important to keep the pairing channel
under control. It shows that the frequently used constant gap
approximation provides unphysical results for the height of fission
barrier. On the contrary, the constant strength approximation shows a
similar functional dependence of the height and the shape of barrier
as the RHB approach. Therefore it is important to treat the pairing
properties in a self-consistent way.

 Seniority forces or zero range forces depend at least on two
parameters, the strength and the cut-off energy. They are usually
adjusted to the experimental pairing properties in the ground state
for a fixed value of the cut-off energy. It is shown that this
procedure leaves room for uncertainties in the fission barriers of up
to 1 MeV.

The extrapolation to superheavy nuclei leads to some uncertainties in
the definition of the pairing strength in the $Z=120$ and $126$
regions of superheavy nuclei. It was shown that 10\% uncertainties in
the pairing strength have drastic consequences for the structure of
these superheavy nuclei.

The DD-ME2 force is the only presently known parameterization of the
RMF Lagrangian which provides a good description of fission barriers
in superheavy nuclei in axially symmetric calculations. Previous and
current studies within covariant density functional theory show that
other RMF forces either considerably underestimate the barriers or do
not leave the room for the triaxiality at the saddle point.

\bigskip

{\leftline{\bf ACKNOWLEDGEMENTS} This work has been supported by the
U.S. Department of Energy under the grant DE-FG02-07ER41459, by the
Bundesministerium f\"{u}r Bildung und Forschung (BMBF), Germany, under
Project 06 MT 246, by the DFG cluster of excellence \textquotedblleft
Origin and Structure of the Universe\textquotedblright\
(www.universe-cluster.de), and by the Hellenic State Scholarship
foundation (IKY).}



\end{document}